\documentclass[sigconf]{acmart}

\renewcommand\footnotetextcopyrightpermission[1]{} 
\acmConference[ICPP '22]{51st International Conference on Parallel Processing}{August 29-September 1, 2022}{Bordeaux, France}

\usepackage{graphicx}
\usepackage{color}

\newtheorem{assumption}{Assumption}

\usepackage{graphicx}
\usepackage{algorithm}
\usepackage[noend]{algpseudocode}
\usepackage{amsmath}
\renewcommand{\algorithmicrequire}{\textbf{Input:}}
\renewcommand{\algorithmicensure}{\textbf{Output:}}
\usepackage{breqn}
\usepackage{subcaption}
\usepackage{multirow}
\usepackage{psfrag}
\usepackage{url}
\usepackage{hyperref}
\usepackage{cleveref}
\crefname{assumption}{assumption}{assumptions}
\usepackage{booktabs}
\usepackage{rotating}
\usepackage{colortbl}
\usepackage{epstopdf}
\usepackage{nag}
\usepackage{microtype}
\usepackage{siunitx}
\usepackage{nicefrac}
\usepackage{lipsum}
\usepackage[english]{babel}
\usepackage[pangram]{blindtext}
\usepackage{tikz}
\usepackage{tabularx}
\usetikzlibrary{fit,positioning,arrows.meta,shapes,arrows}

\begin{document}
%
%
\title[FAIR-BFL: Flexible and Incentive  Redesign for Blockchain-based Federated Learning]{FAIR-BFL: Flexible and Incentive  Redesign \\
for Blockchain-based Federated Learning}
\titlenote{To appear in ICPP '22}

\author{Rongxin Xu}
\affiliation{%
  \institution{Hunan Key Laboratory of Data Science \& Blockchain, Business School,
Hunan University}
  \city{Changsha 410082}
  \country{China}
  }
\email{rongxinxu@hnu.edu.cn}
\orcid{0000-0001-9831-7513}

\author{Shiva Raj Pokhrel}
\authornote{Corresponding Author}
\affiliation{%
  \institution{School of IT,  Deakin University}
  \city{Geelong, VIC 3216}
  \country{Australia}}
\email{shiva.pokhrel@deakin.edu.au}
\orcid{0000-0001-5819-765X}

\author{Qiujun Lan}
\affiliation{%
  \institution{Hunan Key Laboratory of Data Science \& Blockchain, Business School,
Hunan University}
  \city{Changsha 410082}
  \country{China}
  }
\email{lanqiujun@hnu.edu.cn}
\orcid{0000-0001-7523-9487}

\author{Gang Li}
\affiliation{%
  \institution{Centre for Cyber Security Research and Innovation, 
Deakin University}
  \city{Geelong, VIC 3216}
  \country{Australia}}
\email{gang.li@deakin.edu.au}
\orcid{0000-0003-1583-641X}

\renewcommand{\shortauthors}{Rongxin Xu, Shiva Raj Pokhrel, Qiujun Lan, and Gang Li}

\begin{abstract}
	Vanilla  
	\emph{Federated learning} (FL) relies on the centralized global aggregation mechanism 
	and assumes that all clients are honest. 
	This makes it a challenge for FL to alleviate the \emph{single point of failure} 
	and dishonest clients. 
	These impending challenges in the design philosophy of FL
	call for blockchain-based federated learning (BFL) 
	due to the benefits of coupling  FL and blockchain
	(e.g., democracy, incentive, and immutability). 
	However, one problem in vanilla BFL is that its capabilities do 
	not follow adopters' needs in a dynamic fashion. 
	Besides, vanilla BFL relies on unverifiable clients'
	self-reported contributions like data size 
	because checking clients' raw data is not allowed in FL for privacy concerns. 
	We design and evaluate a novel BFL framework, 
	and resolve the identified challenges in vanilla BFL 
	with greater flexibility and incentive mechanism called \emph{FAIR-BFL}. 
	In contrast to existing works, 
	\emph{FAIR-BFL} offers unprecedented flexibility 
	via the modular design, 
	allowing adopters to adjust its capabilities following 
	business demands in a dynamic fashion.
	Our design accounts for BFL's ability to 
	quantify each client's contribution to the global learning process. 
	Such quantification provides a rational metric for 
	distributing the rewards among federated clients 
	and helps discover malicious participants 
	that may poison the global model. 
\end{abstract}


\maketitle


\section{Introduction}\label{sec-intro}

The advent of \emph{federated learning} (FL)~\cite{konecny_federated_2016} 
has ameliorated the shortcomings of the centralized ML techniques,
which were caused by the ever-increasing data scale and model complexity. 
FL addresses the concerns on data ownership and privacy 
by ensuring that 
no raw data leave the distributed end devices (also referred to as clients). 
It successfully employs a single global server 
in a distributed system to collect updates from end devices 
~\cite{pokhrel_federated_2020}. 
FL performs the renewal aggregation
and iteratively distributes new global learning model to the clients. 
However, 
such a FL setup based on centralized server 
suffers from issues such as 
\emph{single point of failure} and instability~\cite{romanFeaturesChallengesSecurity2013}. 
Moreover, 
attacks against distributed training of FL have revealed that 
malicious or compromised clients/central servers 
may upload modified global parameters, 
causing model poisoning, 
because attackers can forge local updates to 
launch inference attacks~\cite{nasr_comprehensive_2019}. 
Therefore, 
the design of a robust FL mechanism 
is essential to 
the stability and security of distributed computing systems.

As a proven decentralized framework, 
blockchain naturally ponders the benefits of merging with FL~\cite{nguyen_federated_2021}, 
including immutability, traceability, and incentive mechanisms, 
let alone the fact that 
both blockchain and FL are inherently distributed by nature. 
Several recent works have been proposed to 
empower FL's robustness, 
intelligence, and privacy-preserving capabilities 
by incorporating blockchain. 
\emph{Blockchain-based Federated Learning} (BFL), 
proposed in \cite{pokhrel_federated_2020}, 
has been considered as
one promising and celebrated approach to 
facilitating distributed computing and learning. 
Notable studies along this line of research include
~\cite{pokhrel_federated_2020,pokhrelFederatedLearningMeets2020,pokhrelDecentralizedFederatedLearning2020,pokhrelBlockchainBringsTrust2021}. 
In BFL, 
local updates and global models can be recorded 
through the blockchain to ensure security, 
and clients would automatically acquire new global parameters 
through a consensus mechanism. 
However, 
BFL requires flexibility 
because adopters' needs are dynamic, 
e.g., when business shrinks, 
adopters may expect to quickly switch from BFL to degraded versions 
(FL or blockchain) for the sake of cost reduction. 
Moreover, 
blockchain rewards those nodes that win the mining competition,
but in BFL, 
we desire to attract potential participants and keep clients 
who make great contributions to global updates. 
Therefore, 
BFL also needs a novel incentive mechanism. 
Unfortunately, 
existing works have not adequately studied 
the flexibility and incentive mechanism in BFL, 
which we refer to as vanilla BFL, 
thus it is difficult for them to move toward practical use.
There exist three more challenges 
in moving vanilla BFL towards flexibility and effective incentive.
\begin{description}
	\item[Tightly coupling blockchain and FL]
	FL has a periodic 
	learning-updating-waiting process 
	while the blockchain keeps running. 
	In vanilla BFL, 
	these two play almost independently, 
	thus posing severe concerns. 
	One prevalent concern is unwanted consequences 
	such as ``\emph{forking is inevitable}''
	\cite{pokhrel_federated_2020}. 
	Ameliorating the impact of forking 
	(as studied in~\cite{cao_when_2019,ma_when_2021}) 
	is non-trivial 
	as it often loses some local updates and 
	adversely impacts global learning. 
	It becomes intractable 
	when local updates recorded in the block 
	can not reflect the actual FL stage 
	and generates empty blocks~\cite{bao_flchain_2019}.
	Therefore, 
	it is crucial to tightly couple both blockchain and FL in BFL, 
	especially to coordinate miners' behavior.

	\item[Defining block's data scope]
	All nodes in the blockchain network have access to 
	the data in the block, 
	so any data which may reveal the privacy 
	must be avoided from being made public. 
	As the block size is limited, 
	vanilla BFL may generate more blocks 
	because it records all the data to complete the same round of learning. 
	Moreover, 
	each block results from a round of mining competition, 
	and large blocks can increase the transfer time. 
	Thus, 
	the delay of vanilla BFL can be high. 
	So the data recorded in the block  
	should be carefully defined to reduce the delay of BFL.
	
	\item[Incentivizing based on contribution]
	Blockchain can provide incentives for FL, 
	whereas it rewards those miners that successfully mine blocks. 
	BFL desires to reward clients 
	who contribute more to the global aggregation 
	to attract potential participants, 
	especially in data-intensive tasks. 
	To this end, 
	a method is needed to help BFL differentiate the client contributions. 
	At the same time, 
	such a method should not rely on clients' self-reported contributions. 
	Otherwise, 
	clients could have good reasons to cheat, 
	and the BFL cannot verify who are dishonest 
	since the limitation in checking the client's raw data. 
	Unfortunately, 
	vanilla BFL mainly relies on the client's self-reporting 
	contributions or checking raw data to determine rewards.
\end{description}

Therefore, 
flexibility and effective incentive should be considered in BFL 
to fully move towards practical use. 
It requires enhancing vanilla BFL with a tightly coupled framework and 
contribution-based incentive mechanism 
to improve performance and security. 
The above challenges motivate us to 
develop new insights in designing the BFL framework.
To this end, 
we propose \emph{FAIR-BFL},
a novel BFL framework 
with \emph{flexible and incentive redesign},
which mitigates the above-mentioned issues in vanilla BFL,
and our main contributions can be summarized as follows:
\begin{enumerate}
	\item We develop new insights in designing blockchain-based federated learning 
	framework by coordinating miners' behavior, 
	recording the desirable global gradients, 
	thus, 
	jointly improving the flexibility and enhancing performance-cum-security.
	
	\item We propose a contribution-based incentive mechanism 
	that supports quantifying each client's contribution 
	with various clustering algorithms, 
	defending against malicious attacks, 
	and selecting high-contributing 
	clients to speed up model convergence.
	
	\item With the incentive mechanism in \emph{FAIR-BFL}, 
	we propose an aggregation method to assign clients' weights 
	based on their contributions, 
	which improves the performance considerably 
	with guaranteed fairness and convergence.
\end{enumerate}

The rest of this paper is organized as follows. 
We start with \Cref{sec-overview}, 
which provides background knowledge on BFL, 
related work, and challenges faced by vanilla BFL. 
Then, 
we propose \emph{FAIR-BFL} in \Cref{sec-modeling} 
and show how it overcomes the challenges with several novel insights.
In \Cref{sec-scaling}, 
we reveal how \emph{FAIR-BFL} provides unprecedented flexibility 
through functional scaling and analyze its performance. 
After that, 
we move to experiments in \Cref{sec-validation} 
to demonstrate the performance, latency, 
and security of \emph{FAIR-BFL}. 
Finally, 
\Cref{sec-conclusions} concludes this work.

\section{Background and Related Work}\label{sec-overview}

Blockchain maintains a distributed ledger to securely record 
conclusive information (called \emph{block}), 
in which the nodes compete for bookkeeping 
in a mining competition 
and reach agreement through a consensus mechanism. 
The newly generated block is broadcasted in the network, 
and those who receive the message will stop their current computation. 

FL employs a distributed learning 
which allows end devices to train their own models locally 
and then aggregates intermediate information 
to provide global insights by using local learnings 
at a central server. 
It aims to solve the problem of 
\emph{data island}~\footnote{Imagine data as a flowing ocean from 
	which some entities collect data and keep it locally 
	rather than sharing it, e.g., financial institutions. 
	Thus, 
	these stagnant data become islands one after another.} 
and benefit from aggregate modelling.
Specifically, 
at the beginning of each communication round, 
the clients update their local models with their data 
and upload the obtained gradients to the central server. 
After that, 
the central server computes the global updates 
by aggregating all received local gradients and 
supplies the global gradients to the clients. 
Finally, 
the clients apply the global gradients to 
update their local models independently. 
Thus, 
FL dynamics evolves from one round to another.

Some notable studies along the lines of BFL include
~\cite{awan_poster_2019,majeed_flchain_2019,lu_blockchain_2020,li_blockchain-based_2021,pokhrel_federated_2020}.
Among them,
\citet{awan_poster_2019} designed a variant of 
the \emph{Paillier} cryptosystem
to support additional homomorphic encryption 
and proxy re-encryption,
so that the privacy of the transmission is protected in the BFL.  
\citet{majeed_flchain_2019} adopted the concept of ``channels'' 
in their BFL framework, 
\emph{FLchain},
to store the gradient of the local model 
in the block of a channel. 
Components such as \emph{Ethereum}, 
extend the capability of \emph{FLchain} 
in executing the global updates. 
\citet{lu_blockchain_2020} incorporated federated learning with 
differential privacy into permissioned blockchain 
to alleviate the centralized trust problem. 
\citet{li_blockchain-based_2021} applied the blockchain to 
store the global model and exchange the local updates, 
thus eliminating the central server and 
resisting privacy attacks from malicious clients and central server,
it also reduces the computation time using committee-based consensus.

The aforementioned works developed the vanilla BFL framework 
for better privacy and data security.
However, 
vanilla BFL design still faces some non-trivial challenges. 
For example, 
the asynchronous nature of blockchain requires in-depth integration 
into the FL's communication rounds mechanism. 
In vanilla BFL frameworks, 
blockchain and federated learning are more like 
two disparate parts that 
are unnecessarily aligned in terms of their working states, 
making it problematic to ensure coherent BFL operations. 
Moreover, 
in most existing vanilla BFL frameworks, 
objective evaluation of the client's contribution 
has been considered as irrelevant, 
while recording all local updates 
in a few of them~\cite{pokhrel_federated_2020,kim_blockchained_2020} 
arose serious privacy leakage concerns. 
None of the aforementioned works 
focus on the \emph{flexibility and incentive mechanism of the BFL by design}.
In this paper, 
we propose \emph{FAIR-BFL} with a modular design, novel insights, 
a fairer aggregation scheme and 
a contribution-based incentive mechanism, 
thus jointly enhancing the performance and the security.

\section{Flexible and Incentive Redesign for BFL}\label{sec-modeling}

In this section, 
we propose \emph{FAIR-BFL} and 
develop the algorithm in detail,
and then we demonstrate 
how to integrate blockchain and FL tightly  
by utilizing their internal working principles. 
\Cref{tab: notations} summarizes all the notations used in this paper.

\begin{table}[htbp]
	\centering
	\caption{Summary of notations}
	\begin{tabular}{ll}
		\toprule
		\multicolumn{2}{c}{Notations in this work} \\
		\midrule
		${C_i}$    & The client $i$ in BFL and FL, or a worker in blockchain. \\
		${S_k}$    & The miner $k$ in BFL and blockchain, or a server in FL. \\
		$\mathcal D$    & The Data set we used \\
		$\lambda$ & The ratio of randomly selected clients in each round \\
		$\eta$ & The learning rate of the model we use \\
		$E$   & The number of epochs of the client's local model \\
		$B$    & The batch size of client's local model \\
		$n$    & The number of clients or workers \\
		$m$    & The number of miners \\
		$\mathcal B$    & Sub-data sets divided by batch size \\
		$w$   & The gradient in FL or BFL \\
		\bottomrule
	\end{tabular}%
	\label{tab: notations}%
\end{table}%

\begin{figure}[htbp]
	\centering
	\includegraphics[width=0.8\linewidth]{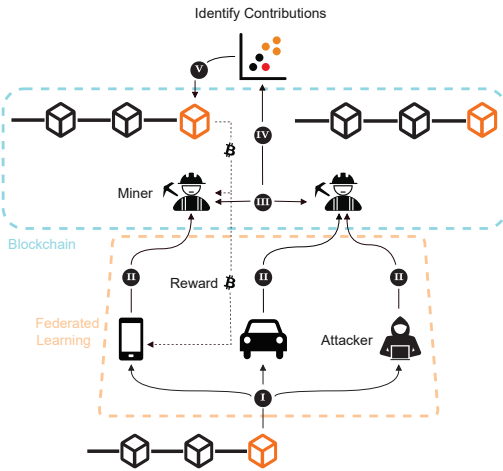}
	\caption{The framework of \emph{FAIR-BFL}}
	\label{fig:bfl}
\end{figure}

\begin{algorithm}[htbp]  
	\caption{\emph{FAIR-BFL} Algorithm}  
	\label{alg:bfl}  
	\begin{algorithmic}[1]  
		\State Initialization: $\{ {C_i}\} _{i = 1}^n,\{ {S_k}\} _{k = 1}^m,\mathcal D,\lambda, \eta, E, B$
		\For{each round $r = 1,2,3...$}
		\State $\{ {C_i}\} _{i = 1}^{\lambda n} \leftarrow Randomly\ select\ \lambda n\ {C_i} \in \{ {C_i}\} _{i = 1}^n$
		\ForAll {${C_i} \in \{ {C_i}\} _{i = 1}^{\lambda n}$}
		\State allocate ${\mathcal D_i}{\sim}\mathcal D$ to ${C_i}$
		\Procedure{\textcolor{blue}{Local model update}}{$C_i$, $\mathcal D_i$, $B$, $E$}
		\State read global gradient ${w_r}$ from the latest block 
		\State $\mathcal B \leftarrow split\ {\mathcal D_i}\ into\ batches\ of\ size\ B$ 
		\For{$each\ epoch\ i\ from\ 1\ to\ E$}
		\For{$each\ batch\ b\ \in\ \mathcal B$}
		\State $w _ { r+1 } ^ { i} \leftarrow w _ { r } ^ { i } - \eta \nabla \ell ( w _ { r } ^ { i } ; b )$
		\EndFor
		\EndFor
		\EndProcedure
		\Procedure{\textcolor{blue}{Upload local gradients}}{$C_i$, $w _ { r+1 } ^ { i}$, $S_k$}
		\State randomly associate ${C_i}$ to ${S_k}$
		\State upload updated gradient $w _ { r+1 } ^ { i}$ to ${S_k}$
		\EndProcedure
		\EndFor
		\ForAll {${S_k} \in \{ {S_k}\} _{k = 1}^m$}
		\Procedure{\textcolor{blue}{Exchange gradients}}{$\{ w_{r+1}^{i}\}$, $S_k$}
		\State $W _ { r+1 } ^ { k }\leftarrow \{ w_{r+1}^{i},i = index\ of\ associate\ clients\}$
		\State broadcast clients updated gradient $W _ { r+1 } ^ { k }$
		\State received updated gradient $W _ { r+1} ^ { v}$ form ${S_v}$
		\For {$w \in W _ { r+1} ^ { v}$}
		\If {$w \notin W _ { r+1 } ^ { k }$}
		\State $W _ { r+1 } ^ { k }$ append $w$
		\EndIf
		\EndFor		
		\EndProcedure
		\Procedure{\textcolor{blue}{Computing Global Updates}}{$W _ { r+1 } ^ { k }$, $S_k$}
		\State $w_{r + 1} \leftarrow \frac{1}{n}\sum\limits_{i = 1}^n {w_{r + 1}^i},w_{r+1}^{i} \in {W_{r+1}^{k}}$ \Comment{Simple Average}
		\State $W _ { r+1 } ^ { k }$ append ${w_{r+1}}$
		\State $Contribution\text{-}based\ Incentive\ Mechanism(W_{r+1}^{k})$ \label{alg:cii-in-alg:bfl}
		\State $Fair\ Aggregation(W_{r+1}^k)$ \Comment{By \Cref{eq:contriAVG}}
		\EndProcedure
		\Procedure{\textcolor{blue}{Block Mining and Consensus}}{$w_{r+1}$, $S_k$}
		\State do $proof\ of\ work$
		\If {hash satisfies target} 
		\State $ Trans. \leftarrow reward\ list$
		\State generate and add $block(Trans., w_{r+1})$
		\State broadcast 
		\EndIf
		\If {received $bloc{k_i}$}
		\State verify $proof\ of\ work$
		\If {hash satisfies target} 
		\State stop current $proof\ of\ work$
		\State blockchain add $bloc{k_i}$
		\EndIf
		\EndIf
		\EndProcedure
		\EndFor
		\EndFor
	\end{algorithmic}  
\end{algorithm} 

A high-level view of \emph{FAIR-BFL} framework 
is shown in \Cref{fig:bfl},
in which the circled number indicates 
the corresponding procedure defined and explained 
in \Cref{alg:bfl} and \Cref{sec-scaling}.
We summarize the entire process of BFL into five procedures 
that interact among different entities: 
i) the client reads the global parameters from the latest block 
	and updates its local model;
ii) the client connects to a miner and uploads its local gradient, 
	please note that some clients may be malicious;
iii) miners exchange gradient sets and start the mining competition; 
iv) the winner identifies the contributions and computes the global updates, 
	and this will help disregard the information forged by the attackers 
	due to low contribution; 
v) the winner packs the global update and reward information 
	into a new block, 
	and then all miners reach an agreement 
	through the consensus mechanism.
More specifically,
the whole process is a holistic approach 
with multiple rounds of communication among $n$ clients
$\{ {C_i}\} _{i = 1}^n$ with a set of $m$ miners 
$\{ {S_k}\} _{k = 1}^m$ 
to handle the blockchain process. 
We then utilize data set $\mathcal D$, 
by assuming that 
the sub dataset ${\mathcal D_i}$ is at the client ${C_i}$ 
before the start of each communication round. 

\subsection{Coupled BFL design}

Vanilla BFL design faces some challenges in flexibility and privacy~\cite{pokhrel_federated_2020}. 
On the one hand, 
the workflows of FL and blockchain are inconsistent 
in the vanilla BFL.
Hence, 
miners will continue to 
compete for excavation without stopping, 
which undoubtedly increases resource consumption. 
For example, 
if a miner does not receive any gradient update
but completes the hash puzzle ahead of other miners, 
it will generate an empty block, 
which does not benefit the FL part. 
However, 
~\cite{chen_revisiting_2017} have shown that 
performing SGD based on communication rounds in FL 
is better than asynchronous methods.
To this end, 
we bring it into BFL to achieve the tight coupling 
between blockchain and FL, 
also alleviate issues such as forking, empty blocks, 
and resource cost.
Note that \Cref{ass1} has been made in~\cite{kim_blockchained_2020}, 
but only to simplify the problems for analysis. 
We have the following \Cref{ass1}. 

\begin{assumption}[Tight coupling]\label{ass1}
	Clients and miners are fully synchronized 
	in every communication rounds. 
\end{assumption} 

On the other hand,
vanilla BFL records every local gradient in the blockchain, 
and workers read the block's information to 
calculate the global updates themselves. 
In this case,
vanilla BFL is a white-box for the attacker, 
malicious nodes can use this information to 
perform privacy attacks and easily track the changes 
in a worker's local gradient to 
launch more severe model inversion attacks~\cite{fredrikson_model_2015}. 
Furthermore,
imagining the BFL applications in large-scale scenarios,
where thousands of local gradients could be waiting for miners to pack. 
However, 
the block size is limited 
due to the communication cost and delay, 
and many local gradients will miss the current block. 
Eventually, 
to calculate the global gradient, 
workers have to wait for a new block to be generated  
until all local gradients have been recorded in the blockchain, 
which undoubtedly increases the latency and 
the communication costs. 
To address the concerns we have the following  \Cref{ass2}.

\begin{assumption}[Bounding block's data scope]\label{ass2}
	Miners pack only the global gradients into blocks. 
	In the end, 
	each block contains only the global gradient of a specific round. 
\end{assumption}  

Observe that \Cref{ass2} is to bound the block's data scope,
also protects the security of \emph{FAIR-BFL} 
and alleviates the transaction queuing 
caused by the limitation of block size in the asynchronous design. 
To the best of our knowledge, 
this is the first attempt to 
use \Cref{ass2} in the BFL design and validate its capability.  

\subsection{Accounting Client's Contribution}\label{sec-cii}

\begin{algorithm}[htbp]  
	\caption{Client's Contribution Identification Algorithm}  
	\renewcommand{\algorithmicrequire}{\textbf{Input:}}
	\renewcommand{\algorithmicensure}{\textbf{Output:}}
	\label{alg:cii}  
	\begin{algorithmic}[1]  
		\Require $W_{r+1}^{k}$, $model\ name$, $Strategy$
		\State $Group\ List \leftarrow Clustering(model\ name, W_{r+1}^{k})$
		\For{$l_i \in Group\ List$} 
		\If{$w_{r+1} \in l_i$} 
		\For{$w^i_{r+1} \in l_i$}
		\State $\theta _{i} \leftarrow Cosine\ Distance(w^i_{r+1},w_{r+1})$
		\State \textbf{Label} $C_i$ as high contribution
		\State \textbf{Append} $\langle C_i, \theta _i/{{\sum\limits_{k = 1}^{\lambda n} {{\theta _k}} }}*base\rangle$ to $reward\ list$
		\EndFor
		\EndIf
		\If{$w_{r+1} \notin l_i$}
		\ForAll{$w^i_{r+1} \in l_i$}
		\State \textbf{Label} $C_i$ as low contribution
		\EndFor
		\EndIf
		\EndFor
		\State $W_{r+1}^k \leftarrow Strategy(reward\ list, W_{r+1}^{k})$
		\Ensure $reward\ list$, $W_{r+1}^k$
	\end{algorithmic}  
\end{algorithm}

\Cref{alg:cii} implements our method to 
identify contributions in \Cref{alg:cii-in-alg:bfl} of \Cref{alg:bfl}.
Various clusters of gradients are found 
by applying a clustering algorithm on $W^k_{r+1}$; 
moreover, 
they imply different contributions. 
Note that any suitable clustering algorithm can be used here as needed, 
However, 
we use \emph{DBSCAN} in experiments by default 
because it is efficient and straightforward.
Those clients belonging to the same cluster as the global gradient 
can be considered a high contribution and be rewarded, 
while those far from the global gradient can be considered 
a low contribution and adopt a predetermined strategy. 
There are two strategies:  
i) keep all gradients;
ii) discard low-contributing local gradients 
	and recalculate the global updates $w_{r+1}$. 
The \emph{cosine} distance $\theta_{i}$ 
(the larger the $\theta$, the farther the distance.)
between its local gradient and the global update 
is calculated as the weight of its contribution 
to the global update for a high contributing client $C_i$. 
We can set a $base$ and multiply it by 
$\theta _i/{{\sum\limits_{k = 1}^{\lambda n} {{\theta _k}} }}$
as the final reward for client $C_i$. 
Key-value pairs 
$\langle C_i, \theta _i/{{\sum\limits_{k = 1}^{\lambda n} {{\theta _k}} }}*base\rangle$ 
represent
the reward information, 
and they are recorded in the $reward\ list$. 
Eventually, 
when a miner generates a new block, 
the reward is distributed according to the reward list and 
appended to the current block as transactions. 
After the blockchain consensus is achieved,  
clients will get these rewards.
We explain the intuition behind \Cref{alg:cii} as follows.
\begin{description}
	\item[Privacy preservation]
	Vanilla BFL requires clients to report their data dimensions 
	for rewards determination. 
	Therefore, 
	clients have sufficient motivation to cheat for more rewards. 
	We cannot recognize this deception 
	because it is impossible to check the actual data set, 
	which violates FL's guidelines. 
	On the contrary, 
	as the intermediate information, 
	the gradients can reflect both the data size and the data quality. 
	Using them to perform \Cref{alg:cii} can provide 
	a more objective assessment and ensure privacy.
	
	\item[Malicious attack resistance]
	Malicious clients may upload fake local gradients 
	to attack the global model. 
	The clustering algorithm can find these fake gradients
	because they are different from the real ones~\cite{nasr_comprehensive_2019}.
	We can employ the discarding strategy to avoid skewing 
	the global model with these spurious gradients,
	ultimately maintaining the security of \emph{FAIR-BFL}.
	
	\item[Clients selection]
	If we adopt the discarding strategy, 
	at the same time, 
	the corresponding workers will no longer participate 
	before the round. 
	In this respect, 
	this approach can also be considered 
	as a new method of clients selection, 
	rather than simply random selection. 
\end{description}

We will thoroughly evaluate our contribution-based 
incentive approach in \Cref{sec-validation}.

\subsection{Fair Aggregation}\label{sec-convergenceproof}
The optimization problem considered by \emph{FAIR-BFL} is
\begin{equation*}
	\min\limits_w\left\{{F(\mathbf{w})} \triangleq \sum_{i=1}^{n} p_{i} F_{i}(\mathbf{w})\right\}, 
\end{equation*}
where $F_{i}(\mathbf{w})$, $p_{i}$ are the local objective 
function and weight of clients $i$, respectively. 
Consider the simple average aggregation, which means $p_{1}=p_{2}=...=p_{i}=\frac{1}{n}$:
\begin{equation*}\label{eq:simpleAVG}
	{w_{r + 1}} \leftarrow \frac{1}{n}\sum\limits_{i = 1}^n {w_{r + 1}^i}
\end{equation*}

This is simple average aggregation that treats all clients' gradients equally and averages them.
However, 
clients may not have same sample sizes. 
Thus, 
simple averaging does not reflect such a contribution difference. 
Instead, 
we use the following method to 
aggregate the global gradients for fairness.

\begin{equation}\label{eq:contriAVG}
	{w_{r + 1}} \leftarrow \frac{1}{\lambda }\sum\limits_{i = 1}^n {{p_i}} w_{r + 1}^i, \text{where } {p_i} = \theta _i/{{\sum\limits_{k = 1}^{\lambda n} {{\theta _k}} }}
\end{equation}

That is, 
we assign aggregation weights 
based on the contribution of clients 
to avoid model skew and improve accuracy.
At the same time, 
it is impractical to 
require all devices to participate in the learning process~\cite{choUnderstandingBiasedClient2022,wangTacklingObjectiveInconsistency2020}, 
so we assume that 
all devices are activated before the communication round begins. 
However, 
only some devices are selected to upload local gradients.

Although we use fairness aggregation and partial participation, 
we can still reveal the stability and convergence dynamics of \emph{FAIR-BFL}.
For tractability, 
we have used the following four well-known assumptions 
in literature~\cite{zhang_communication-efficient_2012,
	li_convergence_2020,stich_local_2019,li_federated_2020}.

\begin{assumption}[L-smooth]\label{ass-smooth}
	Consider 
	${F_i}(w) \triangleq \frac{1}{n}\sum\limits_{i = 1}^n \ell  
	\left( {w;{b_i}} \right)$ 
	and ${F_i}$ is L-smooth, 
	then for all $\mathbf{v}$ and $ \mathbf{w}$,
	\[ F_{i}(\mathbf{v}) \leq F_{i}(\mathbf{w})+(\mathbf{v}- \mathbf{w})^{T} \nabla F_{i}(\mathbf{w})+\frac{L}{2}\|\mathbf{v}-\mathbf{w}\|_{2}^{2}. \]
\end{assumption}

\begin{assumption}[µ-strongly]\label{ass-ustrongly}
	${F_i}$ is u-strongly convex, 
	for all $ \mathbf{v} $ and $ \mathbf{w}$,
	\[F_{i}(\mathbf{v}) \geq F_{i}(\mathbf{w})+(\mathbf{v}- \mathbf{w})^{T} \nabla F_{i}(\mathbf{w})+\frac{\mu}{2}\|\mathbf{v}-\mathbf{w}\|_{2}^{2}.\]
\end{assumption}

\begin{assumption}[bounded variance]\label{ass-gradientbound}
	The variance of stochastic gradients in each client is bounded by: 
	\[ \mathbb{E}\left\|\nabla F_{i}
	\left(\mathbf{w}_{r}^{i}, b_i\right)-\nabla F_{i}
	\left(\mathbf{w}_{r}^{i}\right)
	\right\|^{2} \leq \sigma_{i}^{2} \]
\end{assumption}

\begin{assumption}[bounded stochastic gradient]\label{ass-gbound}
	The expected squared norm of stochastic gradients 
	is uniformly bounded, 
	thus for all $ i=1, \cdots, n $ and $ r=1, \cdots, r-1$, 
	we have 
	\[\mathbb{E}\left\|\nabla F_{i}
	\left(\mathbf{w}_{t}^{i}, b_i\right)
	\right\|^{2} \leq G^{2}\]
\end{assumption}

\Cref{ass-smooth,ass-ustrongly} are essential for analyzing the convergence~\cite{zhang_communication-efficient_2012,stich_local_2019,li_federated_2020}. 
Both \Cref{ass-smooth,ass-ustrongly} mandate requirements on the fundamental properties of the loss function. 
That is, the function does not change too fast (\Cref{ass1}) or too slow (\Cref{ass2}).
\Cref{ass-gradientbound,ass-gbound} 
are made in~\cite{li_convergence_2020}.
They enable us to use $w-w^*$ to approximate $F-F^*$.

\begin{theorem}\label{THEOREM1}
	Given \Cref{ass1,ass2,ass-smooth,ass-ustrongly,ass-gradientbound,ass-gbound} hold, 
	\Cref{alg:bfl} converges as follows 
	\begin{equation}\label{eq-theorem1}
		\mathbb{E}
		\left[F\left(\overline{\mathbf{w}}_{r}\right)
		\right]-F^{*} 
		\leq 
		\frac{\kappa}{\gamma+r}\left(\frac{2(B+C)}{\mu}+
		\frac{\mu(\gamma+1)}{2}\left\|\mathbf{w}_{1}-\mathbf{w}^{*}
		\right\|^{2}\right)
	\end{equation}
	where $ \kappa=\frac{L}{\mu} $, 
	$ \gamma=\max \{8\kappa, E\} $, 
	the learning rate 
	$ \eta_{r}=\frac{2}{\mu(\gamma+r)} $, 
	and $ C=\frac{4}{K} E^{2} G^{2} $.
\end{theorem}

\Cref{eq-theorem1} shows that 
the distance between the actual model $F$ and 
the optimal model $F*$
decreases with increasing communication rounds. 
\emph{FAIR-BFL} can converge regardless of the data distribution 
because we did not make an IID assumption, 
and it establishes the condition that 
guarantees the convergence of \Cref{alg:bfl}. 
The detailed proof of \Cref{THEOREM1} 
is provided in \Cref{proof-theorem1}, 
which is further supported 
by the experimental results in \Cref{sec-validation}. 

\section{Flexibility by Design}\label{sec-scaling}

We re-examined the entire process of vanilla BFL and 
identified the opportunity to achieve flexibility. 
More specifically, 
apart from the necessary work in the preparation phase, 
we divide the remaining part into five procedures,
as shown in \Cref{alg:bfl}. 
Depending on the application's needs, 
these five procedures can be coupled flexibly and dynamically. 
We present these procedures in detail and reveal this flexibility,
and we further determine the possible delays in each link 
to model approximate performance.

\subsection{Local Learning and Update}\label{localupdate}

At the beginning of round $r+1$,
each client reads the global gradient ${w_r}$ 
(if any exists) 
from the last block in the blockchain, 
and updates the local model with ${w_r}$. 
Next, 
the allocated sub-data sets ${\mathcal D_i}$ will be divided 
according to the specified batch size $B$.
For each epoch $i \in \{ 1,2,3, \ldots ,E\}$, 
the client ${C_i}$ obtains the gradient $w _ { r+1 } ^ { i}$ 
of round $r+1$ by performing the SGD,
as shown in \Cref{eq3},
where $\ell$ and $\eta$ are the loss function and 
the learning rate, 
respectively. 

\begin{equation}\label{eq3}
	w_{ r+1 } ^ { i} \leftarrow w _ { r } ^ { i } - 
	\eta \nabla \ell ( w _ { r } ^ { i } ; b ) 
\end{equation}

\Cref{eq3} can be calculated $\frac{{\mathcal D}_i}{B}$ times 
with the specified batch size $B$. 
Therefore, 
the time complexity of \cref{eq3} is 
$\mathcal O(E*\frac{{\mathcal D}_i}{B})$.
Furthermore, 
we define the calculation time of this step 
as the delay $\mathcal T_{local}$.
However, 
please note that 
$E$ and $B$ are set as 
small constants for all clients under normal circumstances, 
so the time complexity of \cref{eq3} is linear $\mathcal O(n)$.
The learning rate $\eta$ is often a more concerning issue, 
as it will significantly affect the performance 
and the convergence rate. 
We will explore the impact of the learning rate on \emph{FAIR-BFL} 
in \cref{sec-lr}.

\subsection{Uploading the gradient for mining}\label{uploading}

After the \texttt{Procedure-I}, 
the client ${C_i}$ will 
get the updated gradient $w_{r+1}^{i}$ of round $r+1$ 
and upload it to the miners. 
There are multiple miners in the network, 
and they will pack gradients into blocks.
In addition, 
BFL miners also need to play a role 
similar to the central server.  
The client does not have to contact all the miners, 
which will increase the communication cost, 
and a better choice is that each client 
only uploads gradients to one miner. 
Here, 
we make the probability of selecting each miner as uniform as possible, 
which is determined based on the specific application scenario. 
Specifically, 
the client ${C_i}$ generates the miner's index $k$ uniformly and randomly, 
then it associates the miner ${S_k}$ and 
uploads the updated gradient $w _ { r+1 } ^ { i}$.

\begin{figure}[htbp]
	\centering
	\includegraphics[width=0.8\linewidth]{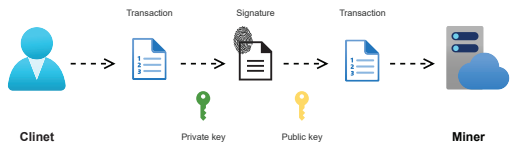}
	\caption{Miners verify transactions through RSA}
	\label{fig:rsa}
\end{figure}

Note that 
it is highly risky to directly use these local gradients  
for subsequent global updates without verifying, 
because malicious clients can easily forge the information and 
launch the gradient attacks~\cite{nasr_comprehensive_2019}.
To avoid this risk, 
we use the RSA encryption algorithm to ensure that 
the identities of both parties are verified.
In the beginning, 
each client is assigned a unique private key according to its ID, 
and the corresponding public key will be held by the miners.
The gradient information received by the miner is signed with the private key, 
so the information can be verified by the public key,
as shown in \Cref{fig:rsa}. 
Further, 
local gradients can be encrypted using RSA to ensure data privacy. 

The procedure will parallelly perform above steps 
for all clients in the current round,
and the time complexity is $\mathcal O(1)$. 
However, 
whether it is selecting miners or uploading gradients, 
the operation itself will be simple. 
Nevertheless, 
the clients are often at the edge of the network, 
and the quality of the channel is difficult to guarantee. 
It may also be subject to other external disturbances, 
where more significant delays are possible. 
For the above considerations, 
we regard the communication time as the main delay in this link 
and record it as $\mathcal T_{up}$. 

\subsection{Exchanging Gradients}\label{minerexchange}

A miner ${S_k}$, 
will get the updated gradient set $\{ w _ { r+1 } ^ { i} \}$ 
from the associated client set $\{ {C_i} \}$,
where $i$ is the index of the clients associated with ${S_k}$. 
In the meantime, 
each miner will broadcast its own gradient set. 
Note that we cancel the queuing here by \Cref{ass1}.
The miner will check whether the received transaction exists 
in the current gradient set $\{w_{r+1}^{i}\}$, 
and if not, 
it will append this transaction. 
In the end, 
all miners have the same gradient set.

Miners will also use the RSA encryption algorithm to 
validate the transactions from other miners to ensure that 
the data has not been tampered with,
as described in \Cref{fig:rsa}. 
The above steps are parallel. 
For each miner, 
it does only three things: 
i) broadcasts the gradient set owned. 
ii) receives the gradient sets from other miners. 
iii) adds the local gradients which it does not own. 
That means, 
the time complexity of the current procedure is $\mathcal O(m)$, 
and we denote the time required from the start 
to the moment when 
all miners have the same gradient set as $\mathcal T_{ex}$.  
Normally, 
the number of miners will be scarce, 
so it is easy to ensure good communications among them, 
which is also the need of the practical application. 
Under such circumstances, 
$\mathcal T_{ex}$ is insignificant.

\subsection{Computing Global Updates}\label{globalupdate}

So far, 
every miner will have all the local updates in this round. 
In order to obtain the global gradient ${w_{r+1}}$, 
they only need to perform fairness aggregation by \Cref{eq:contriAVG}.
So that the clients can initialize the model parameters 
in the $r + 1$ round.

After that, 
to evaluate each client's contribution in this round, 
the global gradient ${w_{r+1}}$ is appended to 
the current local update set $W _ { r+1 } ^ { k }$, 
then we perform \Cref{alg:cii} on $W _ { r+1 } ^ { k }$ 
to identify client contributions and issue rewards.

The current procedure only needs to compute the global gradient 
using \Cref{eq:contriAVG} and then perform \Cref{alg:cii}, 
so the time complexity depends on the clustering algorithm, 
represented as $\mathcal O(clustering)$. 
Also, 
we denote the time cost as the delay $\mathcal T_{gl}$.

\subsection{Block Mining and Consensus}\label{mining}

Once the global gradient calculation is completed, 
all miners will immediately enter the mining competition. 
Specifically, 
the miner will continuously change the nonce in the block header, 
and then calculate whether the block's hash meets the $Target$ by SHA256. 
The whole process can be explained as \Cref{eq5}, 
where $Targe{t_1}$ is a large constant, 
representing the maximum mining difficulty.
Note that 
$Target$ is the same for all miners, 
and mining difficulty will be specified 
before the algorithm starts. 
Therefore, 
the probability that a miner obtains the right to generate blocks 
will depend on the speed of the hash calculation.

\begin{equation}\label{eq5}
	H ( \ nonce + \ Block ) < \ Target\  = \frac{{Targe{t_1}}}{{difficulty}}
\end{equation}

If a miner gets the solution of \Cref{eq5} ahead of other miners, 
it will immediately pack the global gradient ${w_{r+1}}$ 
and reward information into a new block, 
and then broadcast this new block.
After receiving this new block, 
other miners will immediately stop the current hash calculation, 
and append the new block to their blockchain copies, 
once the validity of the new block is verified.
Then, 
it will enter the next communication round.
Again, 
please recall \Cref{ass2}, 
with this setting, 
at the end of a communication round, 
the blockchain will only generate one block, 
and the blockchain copies of all miners will be the same, 
which means that we avoid the blockchain forking,
thus there is no need to resolve ledger conflicts 
while reducing the risk of gradient abandonment and consensus delay.
According to \Cref{eq5}, 
the hash value will be calculated several times
until $Target$  is met, 
and the time of the hash calculation is related to the length of the string.
Based on above discussion,
the time complexity is $\mathcal O(n)$.
We record the time cost here as $\mathcal T_{bl}$, 
which can be more significant compared with others.

\begin{figure}[htbp]
	\centering
	\includegraphics[width=\linewidth]{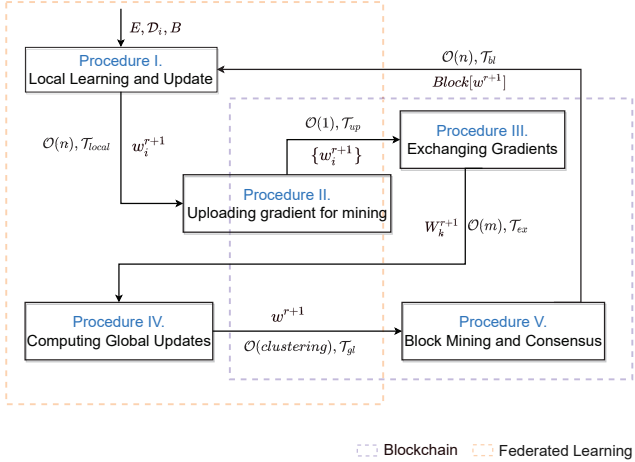}
	\caption{The Coupling structure and complexity of \emph{FAIR-BFL}}
	\label{fig:delay}
\end{figure}

\subsection{Approximate Performance of \emph{FAIR-BFL}}\label{sec-summary}

As shown \Cref{fig:delay}, 
the aforementioned five procedures 
are tightly coupled to form \emph{FAIR-BFL}. 

\begin{description}
	\item[Flexibility Guarantee] 
	If we remove the \texttt{Procedure-I} and \texttt{Procedure-IV},  
	then \emph{FAIR-BFL} boils down to a pure blockchain algorithm 
	(see dashed purple rectangle, \Cref{fig:delay}). 
	On the contrary, 
	if we remove \texttt{Procedure-III} and \texttt{Procedure-V}, 
	it will be equivalent to the pure FL algorithm 
	(see dashed orange rectangle, \Cref{fig:delay}). 
	This scale back functionality by design 
	enables us to easily compare the performance 
	and delay of those three approaches 
	under the same setup using the same data set 
	for their comparison.  
	Moreover, 
	we develop analytic model to quantify this flexibility 
	and analyse the delay of the system \Cref{fig:delay}. 
	
	\item[Approximate Performance Analysis]
	The interactions in \Cref{fig:delay} are explained as follows.
	\texttt{Procedure-I} receives the initial parameters and data set, 
	executes in parallel on each client, 
	and after a delay $\mathcal T_{local}$ 
	it eventually returns the local gradient $w _ { r+1 } ^ { i}$ 
	for a particular client. 
	\texttt{Procedure-II} runs on each miner, 
	receives $w _ { r+1 } ^ { i}$ from the associated client, 
	and spends $\mathcal T_{up}$ time to 
	return the gradient set $\{ w_{r+1}^{i}\}$. 
	After that,
	\texttt{Procedure-III} receives the gradient set $\{ w_{r+1}^{i}\}$ of all miners 
	and then waits for $\mathcal T_{ex}$ time to 
	get the complete local gradient set $W _ { r+1 } ^ { k }$. 
	\texttt{Procedure-IV} uses $W _ { r+1 } ^ { k }$ to 
	calculate the global update $w^{ r+1 }$ of this round, 
	and the time $\mathcal T_{gl}$ consumed 
	depends on the clustering algorithm used. 
	\texttt{Procedure-V} packs the global gradient $w^{ r+1 }$ and 
	generates $Block(Trans., w_{r+1})$, 
	where the above discussion has determined that 
	the delay here is $\mathcal T_{bl}$. 
	Therefore, 
	the overall complexity of \emph{FAIR-BFL} is close to $\mathcal O(n)$, 
	while with $n$ workers and $m$ miners, 
	the overall delay is ${T_{(n,m)}} = {\mathcal{T}_{local}} + 
	{\mathcal{T}_{up}} + {\mathcal{T}_{ex}} + 
	{\mathcal{T}_{gl}} + {\mathcal{T}_{bl}}$, 
	which is compatible with vanilla BFL, 
	so it can quickly learn from \cite{pokhrel_federated_2020} 
	to optimize the block arrival rate to obtain the best delay. 
\end{description}

\section{Evaluation and Discussion}\label{sec-validation}

In this section, 
we conducted a series of experiments to 
comprehensively evaluate the performance of \emph{FAIR-BFL} on real data set,  
Then we reported the changes in performance and delay 
under various conditions by adjusting parameters. 
At last, 
some novel insights such as the trade-off 
between performance and latency are presented.

\subsection{Experimental setup}\label{sec-setup}

Our baseline methods for comparison include the Blockchain, 
\emph{FedAvg}~\cite{mcmahan_communication-efficient_2017}, 
and the state-of-the-art FL algorithm \emph{FedProx}~\cite{li_federated_2020}. 
Then,
we compare the performance of \emph{FAIR-BFL} and baselines
on the benchmark data set \emph{MNIST}.
The metrics for comparison are 
the average delay and the average performance. 
We calculate the average delay 
by $\sum\limits_{i = 1}^r {{d_i}/r}$, 
and the average accuracy by $\sum\limits_{i = 1}^n {ac{c_i}/n}$,
where $d_i$ represents the delay of the communication round $i$, 
$ac{c_i}$ is the verification accuracy of client $C_i$ 
in a communication round.

By default, 
we assign data to clients following the non-IID dynamics, 
and we set $n=100$ and $m=2$, 
$\eta=0.01$, $E=5$, 
and $B=10$, 
respectively. 

\subsection{Performance Impact}\label{sec-delay}

For all experiments, 
We consider the model as converged 
when the accuracy in change is within $0.5\%$ 
for $5$ consecutive communication rounds,
and perform $100$ communication rounds by default.

\subsubsection{General analysis of latency and performance}
\label{sec-general}

\begin{figure}[htbp]
	\centering
	\begin{subfigure}{0.48\linewidth}
		\includegraphics[width=\linewidth]{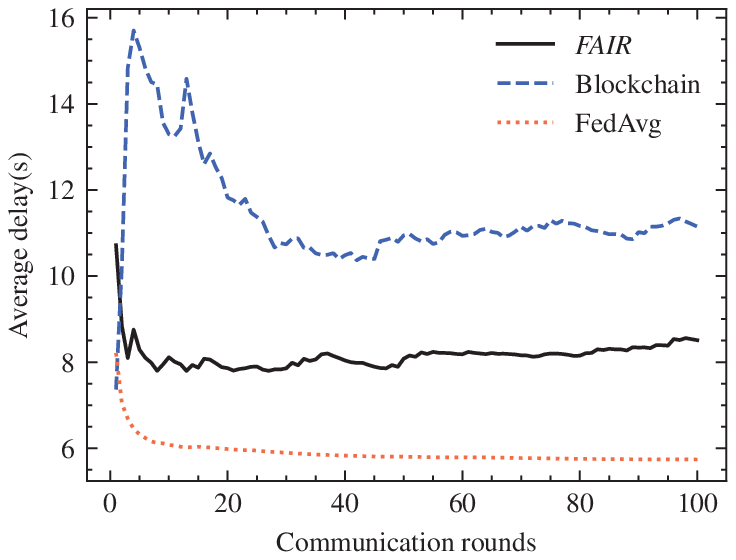}
		\caption{Delay comparison}
		\label{fig:general-delay}
	\end{subfigure}
	\begin{subfigure}{0.48\linewidth}
		\includegraphics[width=\linewidth]{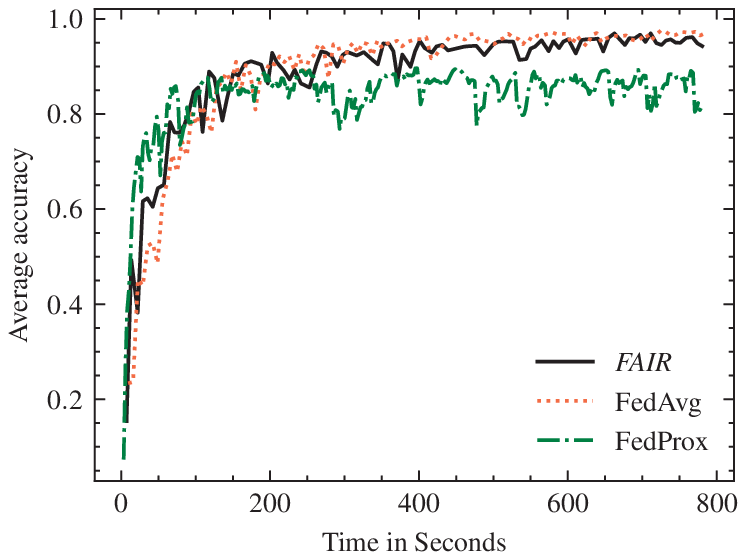}
		\caption{Accuracy comparison}
		\label{fig:general-acc}
	\end{subfigure}
	\caption{General comparison of \emph{FAIR-BFL} and baselines}
	\label{fig:general}
\end{figure}

\Cref{fig:general} shows the simulation results 
of the general delay and performance. 
As shown in \Cref{fig:general-delay}, 
the average delay of \emph{FAIR-BFL} is 
between blockchain and \emph{FedAvg},
rather than above the blockchain.
This implies that 
\Cref{ass1,ass2} can effectively reduce the BFL delay.
In addition, 
from \Cref{fig:general-acc}, 
\emph{FAIR-BFL} has almost the same model performance as the \emph{FedAvg}.
\emph{FedProx} has a lower accuracy than \emph{FAIR-BFL}, 
and the accuracy still fluctuates after the model converges, 
which is because it uses the inexact solution to speed up the convergence. 

\subsubsection{Impact of the Learning rate}\label{sec-lr}

\begin{figure}[htbp]
	\centering
	\begin{subfigure}{0.48\linewidth}
		\includegraphics[width=\linewidth]{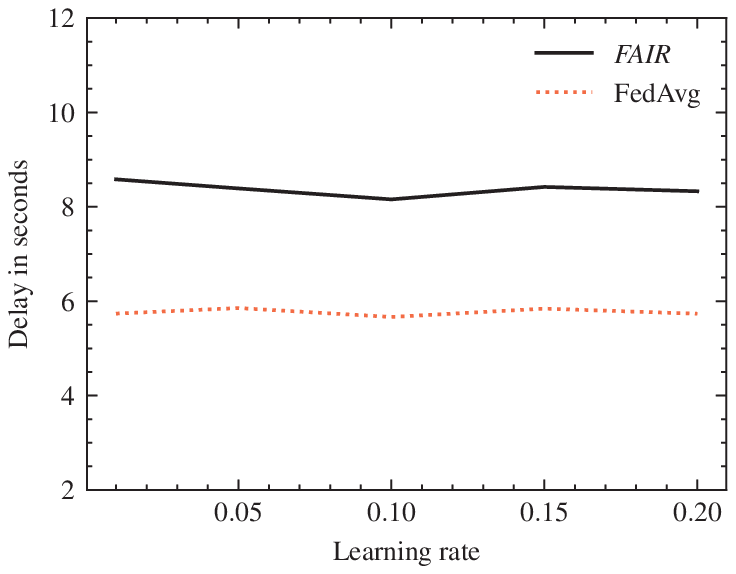}
		\caption{Average delay changes}
		\label{fig:lr-delay}
	\end{subfigure}
	\begin{subfigure}{0.48\linewidth}
		\includegraphics[width=\linewidth]{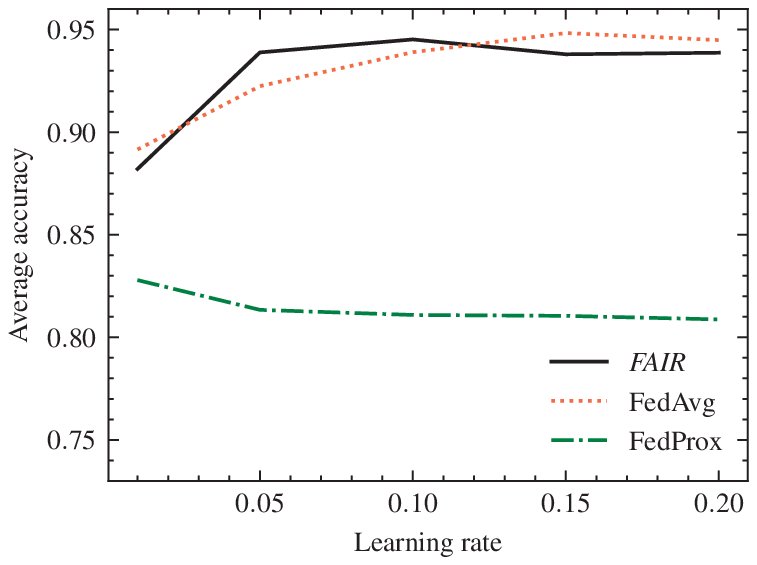}
		\caption{Accuracy changes}
		\label{fig:lr-acc}
	\end{subfigure}
	\caption{Performance and delay under various learning rates}
	\label{fig:lr}
\end{figure}

We conducted multiple experiments, 
where $\eta \in \left[ {0.01,0.05,0.10,0.15,0.20} \right]$.
The result is shown in \Cref{fig:lr}.
From \Cref{fig:lr-delay}, 
we can see that 
the effect of the learning rate on the average delay 
for \emph{FAIR-BFL} and \emph{FedAvg} is negligible, 
and we attribute it to the distributed (parallel) learning method. 
It is interesting to note that 
\emph{FedProx} peaks at the beginning, 
which implies that it may need a larger $\eta$.
Although there is no obvious impact on the delay, 
the accuracy is very different,
as shown in \Cref{fig:lr-acc}. 
For \emph{FAIR-BFL} and \emph{FedAvg}, 
there is an optimal $\eta$ 
such that the average accuracy is the highest. 
For \emph{FedProx}, 
the learning rate does not significantly 
affect the average accuracy.
To this end, 
we have the following insights. 

\textbf{Insight 1:} 
Due to the benefits of distributed learning, 
we set the best learning rate in BFL to 
ensure the model performance. 
The delay overhead for this is acceptable.

\subsubsection{Impact of the number of workers}\label{sec-miners}

\begin{figure}[htbp]
	\centering
	\begin{subfigure}{0.48\linewidth}
		\includegraphics[width=\linewidth]{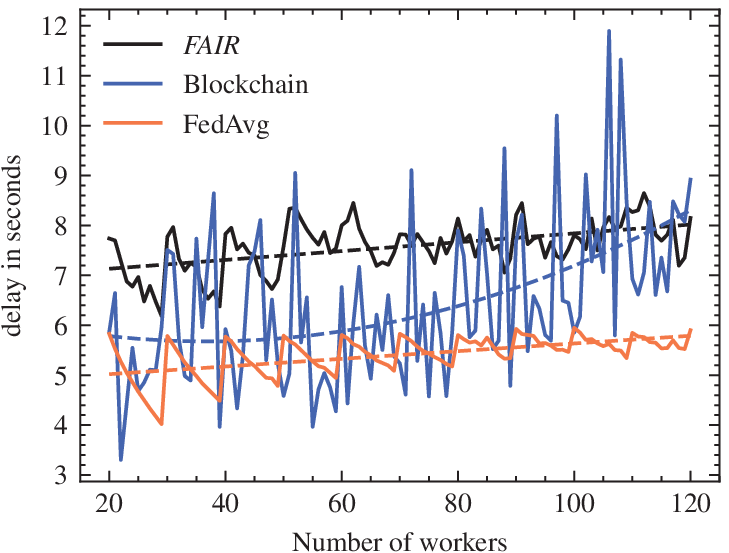}
		\caption{Workers}
		\label{fig:worker-delay}
	\end{subfigure}
	\begin{subfigure}{0.48\linewidth}
		\includegraphics[width=\linewidth]{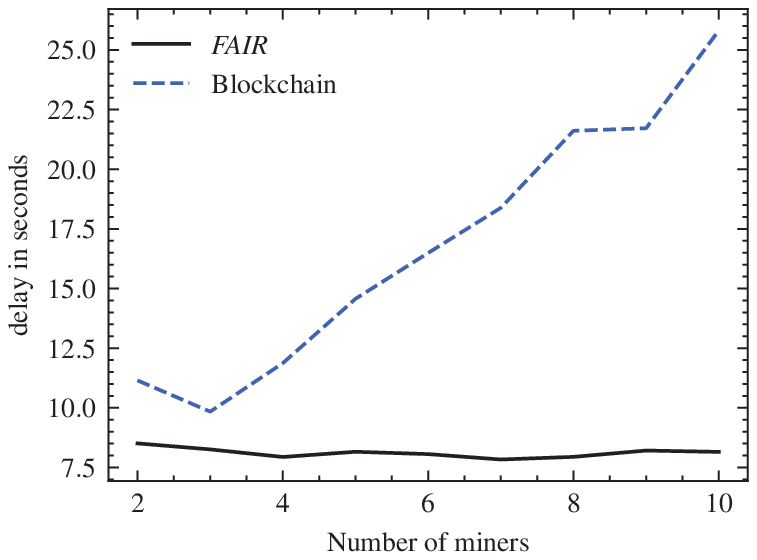}
		\caption{Miners}
		\label{fig:miner-delay}
	\end{subfigure}
	\caption{Average delay changes with the number of workers and miners}
	\label{fig:worker}
\end{figure}

The increase in the number of workers will lead to 
an increase in transactions, 
thus impacting the delay. 
\Cref{fig:worker-delay} shows the delay changes in this case. 
We can see that 
with the increase in the number of workers $n$, 
the delay of blockchain increases,
and this is because the total number of new transactions is rising, 
while the block size is fixed. 
When a block cannot contain all transactions, 
transaction queuing will occur,
which is regarded as a scalability issue~\cite{zhang_towards_2018} in blockchain. 
Please note that 
when the number of new transactions is much smaller than the block size, 
the delay caused by the increase of the clients will be minimal,  
such as the curve in interval $n \in [20,100]$;
when the total size of new transactions crosses the block size ($n\geq100$), 
the delay caused by queuing will become more apparent, 
eventually making the delay of blockchain 
greater than the delay of \emph{FAIR-BFL}, 
which is consistent with the result of \Cref{sec-general}.
On the contrary, 
\emph{FAIR-BFL} achieves a delay similar to \emph{FedAvg}, 
It is almost unaffected by the number of clients. 
Thanks to \Cref{ass1,ass2}, 
no matter how many clients there are, 
there will be no queuing,
because each block only contains the global gradient of the current round. 

\textbf{Insight 2:} 
The block size will significantly affect the delay in large-scale scenarios. 
\Cref{ass1,ass2} provide an effective way to solve this problem.

\subsubsection{Impact of the number of miners}\label{sec-workers}

In contrast to \Cref{fig:worker-delay}, 
we set the number of clients to $100$ 
and increase the number of miners 
to fully observe the impact of this change 
on delay in \Cref{fig:miner-delay}. 
It can be seen that in blockchain, 
the delay increases approximately exponentially 
as the number of miners increases. 
Because when more and more nodes participate in the mining competition, 
the probability of forking will significantly increase, 
which will take more time to merge conflicts. 
For \emph{FAIR-BFL}, 
this issue is avoided due to \Cref{ass1,ass2}. 
Therefore, 
the increase in the number of miners does not significantly increase the delay.

\textbf{Insight 3:} 
Too many miners may cause delay, 
so the number of miners should be set appropriately. 
In BFL, 
we can alleviate this issue by \Cref{ass1,ass2}.

\subsection{Cost-effectiveness}\label{sec-economy}

Here, 
we observe how \Cref{alg:cii} with discarding strategy affects the delay, 
the accuracy 
and the convergence rate of \emph{FAIR-BFL}. 
We use \emph{DBSCAN} as the default clustering algorithm.
Note that \emph{FedProx} also drops clients to 
improve both the convergence rate and the model accuracy. 
However, 
\emph{FedProx} avoids the global model skew by discarding stragglers, 
while we discard the low-contributing clients 
implied by the clustering algorithm. 
To demonstrate the effectiveness of our contribution-based incentive mechanism, 
we set the $drop\text{_}percent$ of \emph{FedProx} to 
$0.02$ as a new baseline for comparison.

\begin{figure}[htbp]
	\centering
	\begin{subfigure}{0.48\linewidth}
		\includegraphics[width=\linewidth]{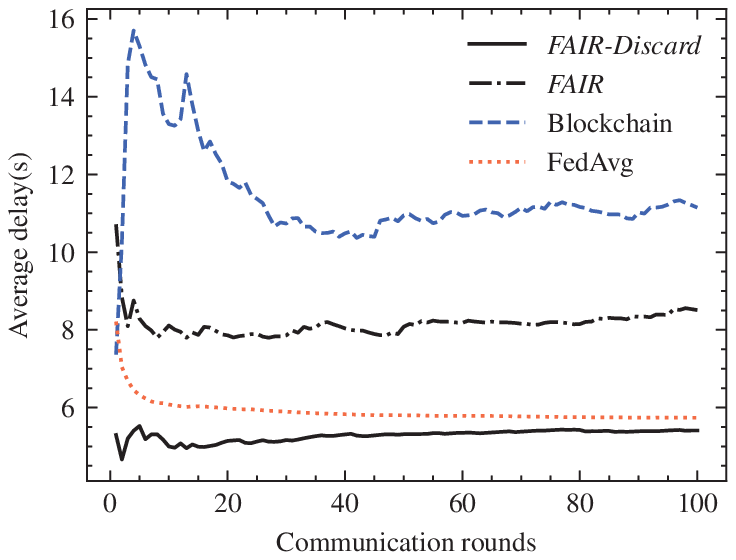}
		\caption{Delay}
		\label{fig:fast-delay}
	\end{subfigure}
	\begin{subfigure}{0.48\linewidth}
		\includegraphics[width=\linewidth]{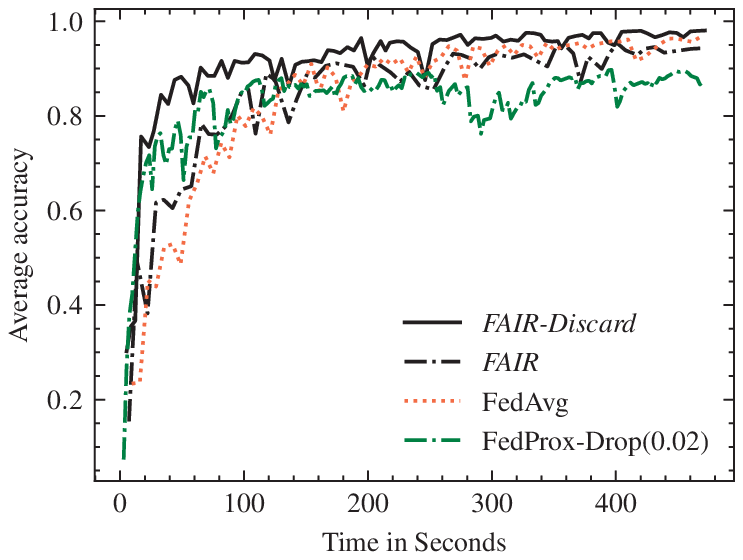}
		\caption{Accuracy}
		\label{fig:fast-acc}
	\end{subfigure}
	\caption{\emph{FAIR-BFL} is faster without reducing accuracy}
	\label{fig:fast}
\end{figure}

Then in \Cref{fig:fast}, 
we observe the difference in accuracy and 
delay with and without the discarding strategy, respectively. 
It can be seen from \Cref{fig:fast-delay} that
the discarding strategy significantly reduces the average delay, 
even lower than that of \emph{FedAvg}.
The reason is that 
those workers with lower contributions no longer participate 
in the current communication round, 
which means fewer workers and local gradients.
On the one hand, 
fewer workers reduce the time cost 
for local updates and upload gradients and 
reduce the total number of communications, 
thus reducing costs. 
On the other hand, 
fewer local gradients accelerate the updates exchange and global aggregation, 
thus reduces the size of data packets in the network,  
to save the traffic and reduce the channel delays.

More importantly, 
the model converges better and faster 
so that the \emph{FAIR-BFL} with discarding strategy in \Cref{fig:fast-acc} 
lies above the \emph{FedAvg} and the original \emph{FAIR}
and reaches the convergence point 
between $250$ and $300$ seconds.
Although \emph{FedProx} also converges better initially, 
its accuracy stabilizes around $84\%$, 
which is lower than \emph{FAIR-BFL}.
The reason is that 
the low-contributing clients no longer 
participate in global aggregation, 
thus reducing the noise from low-quality data, 
effectively preventing the global model from falling into local optimal points,
and improving the accuracy.
As discussed above and shown in \Cref{fig:fast-delay}, 
the discarding strategy significantly reduces the average delay, 
and further reduces the time to reach the convergence.
In conclusion, 
above results confirm that 
\emph{FAIR-BFL} is more economic and faster.  

\textbf{Insight 4:} 
Use the discarding strategy in large-scale scenarios to 
speed up model convergence 
and reduce the cost of communications and traffic.

\begin{table}[htbp]
	\centering
	\caption{Detecting malicious attacks using our contribution-based incentive mechanism}
	\resizebox{\linewidth}{!}{
		\begin{tabular}{lllll}
			\toprule
			\textbf{Distribution} & \textbf{Round} & \textbf{Attacker Index} & \textbf{Drop Index} & \textbf{Detection Rate} \\
			\midrule
			\multirow{9}[1]{*}{\textbf{Non-IID}} & 1 & [3, 7] & [2, 4, 5, 6] & 0\% \\
			& 2 & [3, 6, 2] & [2, 6] & 66.66\% \\
			& 3 & [6, 4, 7] & [4, 6] & 66.66\% \\
			& 4 & [1, 6, 0] & [6] & 33\% \\
			& 5 & [2, 8, 0] & [0, 8] & 66.66\% \\
			& 6 & [7, 0] & [0, 7] & 100\% \\
			& 7 & [0] & [0] & 100\% \\
			& 8 & [3, 9] & [3] & 50\% \\
			& 9 & [6, 0, 8] & [0, 8]& 66.66\% \\
			& 10 & [6, 5]& [5, 6]& 100\% \\
			\midrule
			\multicolumn{4}{l}{\textbf{Average Detection Rate}} 
			& \textbf{64.96\%} \\
			\midrule
			\multirow{10}[1]{*}{\textbf{IID}}& 1 & [0, 6, 1]& [0, 1]& 66.66\% \\
			& 2 & [0, 3, 6]& [3, 6, 8]& 66.66\% \\
			& 3 & [9]& [9]& 100\% \\
			& 4 & [2]& [2]& 100\% \\
			& 5 & [6, 3, 1]& [1, 3]& 66.66\% \\
			& 6 & [5, 9]& [5]& 50\% \\
			& 7 & [3]& [3]& 100\% \\
			& 8 & [7, 0]& [7]& 50\% \\
			& 9 & [1, 7, 2]& [1, 7]& 66.66\% \\
			& 10 & [9]& [9]& 100\% \\
			\midrule
			\multicolumn{4}{l}{\textbf{Average Detection Rate}} & \textbf{75\%} \\
			\bottomrule
		\end{tabular}%
	}
	\label{tab: security}%
\end{table}%

\subsection{Security by design}\label{sec-security}

In \Cref{sec-economy},
we have shown that 
implementing \Cref{alg:cii} with a discarding strategy 
for client selection is effective. 
Here, 
we will demonstrate its security.
We set malicious nodes, 
which modify the actual local gradients to skew the global model.  
At the same time, 
\emph{DBSCAN} is also adopted to find the difference in contribution.
There are $10$ indexed clients, 
and in each communication round, 
randomly designate $1$ to $3$ clients as malicious nodes, 
and $10$ rounds are executed in total, 
as shown in \Cref{tab: security}.

We can see that 
when there are few malicious nodes 
($1$ malicious node in this experiment), 
the detection rate almost always reaches $100\%$,
and \emph{FAIR-BFL} identifies 
the forged gradients as with the low contributions,  
e.g., in the communication round $7$. 
It means that 
with the vast majority of nodes remaining honest, 
the behavior of the malicious nodes is evident,
because the modified local gradients are distant from the normal ones. 
As the number of malicious nodes increase, 
some forged gradients successfully cheat this mechanism, 
so the detection rate decreases,
because anomalies that are obvious enough may mask those 
that are not obvious. 
Even so, 
the detection rate is maintained at an optimistic level, 
for example, 
in round $9$. 
We also found that 
the average detection rate is higher in the case of IID, 
which is attributed to the fact that
a good distribution of data makes the normal gradients 
more spatially concentrated and therefore easier to discover anomalies. 
Interestingly, 
in general, 
the detection rate increases as the model converges. 
The reason is that 
as the model converges, 
individual local gradients are getting similar. 
The results and the above discussion indicate that 
\emph{FAIR-BFL} can resist malicious attacks 
to the greatest extent even in the case of non-IID.

Let's recall the major design aspects 
considered in \emph{FAIR-BFL} to ensure security: 
i) we use the RSA algorithm to sign the local gradient to avoid modification 
	during the upload process (see \Cref{fig:rsa}). 
ii) the data recorded on the blockchain is immutable; 
iii) we use \Cref{alg:cii} to 
	reveal the contribution differences among nodes 
	and discard low contributing local gradients 
	(forged gradients) 
	to resist malicious attacks; 
iv) we do not record any local gradients in the blockchain, 
	so all nodes cannot observe and exploit this information 
	(see \Cref{ass2}). 
Thus, 
\emph{FAIR-BFL} provides the privacy and security guarantee by design
for the whole system dynamics.

\section{Conclusion} \label{sec-conclusions}

We present a new research problem 
and develop valuable insights 
toward modelling blockchain-based federated learning. 
\emph{FAIR-BFL} 
comes with a modular design that 
can dynamically scale functions according to the adopter's needs,
thus providing unprecedented flexibility.
Moreover, 
we provably alleviate
the impending challenges of the vanilla BFL 
in terms of adjusting delay, performance, and aggregation 
by tight coupling blockchain and FL and fairness aggregation.
This is one of the first attempts to
redefine the block's data scope in BFL in order to
prevent clients from observing others' gradients, 
thus enhancing privacy and security. 
More importantly, 
\emph{FAIR-BFL} motivates all clients to contribute actively 
by identifying the client's contribution and issuing uneven rewards. 
Our experimental results show that 
\emph{FAIR-BFL}
can achieve desirable performance beyond the capacity of existing approaches. 
Furthermore, 
\emph{FAIR-BFL} that employs the discarding strategy can naturally reap the benefits of 
privacy, malicious attack resistance, and client selection.

\begin{acks}

This research is supported by the National Natural Science Fund of China (Project No. 71871090), and Hunan Provincial Science \& Technology Innovation Leading Project (2020GK2005). 

\end{acks}

\appendix
\section{Proof of THEOREM \ref{THEOREM1}}\label{proof-theorem1}
As mentioned earlier, $ \mathbf{w}_{r}^{i} $ is the local gradient of the client $C_i$ at communication round $r$.
For tracking the learning process at each local epoch, we use the notion of global epoch and $ \mathcal{I}_{E}=\{n E \mid n= $ $1,2, \cdots n \}$.
For convenience, we denote by $ \mathcal{C}_{r} $ is the most recent set of clients (with size $K$) selected in the communication round $r$.
Note that $ \mathcal{C}_{r} $ results from the random selection, and ergodicity comes from the stochastic gradients.
So we slightly abuse the notation $ \mathbb{E}_{\mathcal{C}_{r}}(\cdot) $,
taking the expectation means that we eliminate the former kind of randomness.

With $\overline{\mathbf{w}}_{r+1}$ denoting the global gradient at round $r+1$ and $\overline{\mathbf{v}}_{r+1}$
tracking the weighted average of all client-side gradients,
we can see that $\overline{\mathbf{w}}_{r+1}$ is unbiased.
In particular, $\overline{\mathbf{w}}_{r+1} = \overline{\mathbf{v}}_{r+1}$. Now,
we formulate the following \Cref{LEMMA1} to bound the variance of $\overline{\mathbf{w}}_{r}$.
\begin{lemma}\label{LEMMA1}
	For $ r+1\in \mathcal{I}_{E} $, if $ \eta_{r} $ is non-increasing and
	$ \eta_{r} \leq2\eta_{r+E} $ for all $ r \geq0$, then the expected difference between
	$ \overline{\mathbf{v}}_{r+1} $ and $ \overline{\mathrm{w}}_{r+1} $ is
	\begin{equation*}
		\mathbb{E}_{\mathcal{C}_{r}}\left\|\overline{\mathbf{v}}_{r+1}-\overline{\mathbf{w}}_{r+1}\right\|^{2} \leq \frac{4}{K} \eta_{r}^{2} E^{2} G^{2}
	\end{equation*}
\end{lemma}

We first provide the proof of \Cref{LEMMA1} which builds the foundation for proving \Cref{THEOREM1}.

\begin{proof}
	Taking expectation over $ \mathcal{C}_{r+1} $, we have
	\begin{equation*}
		\begin{aligned}
			\mathbb{E}_{\mathcal{C}_{r}}\left\|\overline{\mathbf{w}}_{r+1}-\overline{\mathbf{v}}_{r+1}\right\|^{2} & = \mathbb{E}_{\mathcal{C}_{r}} \frac{1}{K^{2}} \sum_{l=1}^{K}\left\|\mathbf{v}_{r+1}^{i_{l}}-\overline{\mathbf{v}}_{r+1}\right\|^{2} \\ &=\frac{1}{K} \sum_{k=1}^{n} p_{i}\left\|\mathbf{v}_{r+1}^{k}-\overline{\mathbf{v}}_{r+1}\right\|^{2}
		\end{aligned}
	\end{equation*}
	where the first equality follows from $ \mathbf{v}_{r+1}^{i_{l}} $ are independent and unbiased. Since $ r+1\in \mathcal{I}_{E} $, the time $ r_{0}=r-E+1\in \mathcal{I}_{E} $ is the underlying communication time, which implies
	$ \left\{\mathrm{w}_{r_{0}}^{k}\right\}_{k=1}^{n} $ are identical. Furthermore,
	\begin{equation*}
		\begin{aligned} \sum_{i=1}^{n} p_{i}\left\|\mathbf{v}_{r+1}^{i}-\overline{\mathbf{v}}_{r+1}\right\|^{2} &=\sum_{i=1}^{n} p_{i}\left\|\left(\mathbf{v}_{r+1}^{i}-\overline{\mathbf{w}}_{r_{0}}\right)-\left(\overline{\mathbf{v}}_{r+1}-\overline{\mathbf{w}}_{r_{0}}\right)\right\|^{2} \\ & \leq \sum_{i=1}^{n} p_{i}\left\|\mathbf{v}_{r+1}^{i}-\overline{\mathbf{w}}_{r_{0}}\right\|^{2} \end{aligned}
	\end{equation*}
	where the last inequality results from \[ \sum_{i=1}^{n} p_{i}\left(\mathbf{v}_{r+1}^{i}-\overline{\mathbf{w}}_{r_{0}}\right)=\overline{\mathbf{v}}_{r+1}-\overline{\mathbf{w}}_{t_{0}} \]
	and \[ \mathbb{E}\|x-\mathbb{E} x\|^{2} \leq  \mathbb{E}\|x\|^{2}. \] Therefore
	\begin{equation*}
		\begin{aligned} \mathbb{E}_{\mathcal{C}_{r}}\left\|\overline{\mathbf{w}}_{r+1}-\overline{\mathbf{v}}_{r+1}\right\|^{2} & \leq \frac{1}{K} \sum_{i=1}^{n} p_{i} \mathbb{E}\left\|\mathrm{v}_{r+1}^{i}-\overline{\mathbf{w}}_{r_{0}}\right\|^{2} \\ & \leq \frac{1}{K} \sum_{i=1}^{n} p_{i} \mathbb{E}\left\|\mathrm{v}_{r+1}^{i}-\mathrm{w}_{r_{0}}^{i}\right\|^{2} \\ & \leq \frac{1}{K} \sum_{i=1}^{n} p_{i} E \sum_{t=r_{0}}^{r} \mathbb{E}\left\|\eta_{r} \nabla F_{i}\left(\mathrm{w}_{r}^{i}, b\right)\right\|^{2} \\ & \leq \frac{1}{K} E^{2} \eta_{r_{0}}^{2} G^{2} \leq \frac{4}{K} \eta_{r}^{2} E^{2} G^{2} \end{aligned}
	\end{equation*}
\end{proof}
As shown in \Cref{globalupdate} we adopt the partial aggregation and therefore the temporal evolution follow
\begin{equation*}
	\begin{aligned}
		\mathbf{v}_{r+1}^{i} & = w_r^i - {\eta _i}\nabla F_i \left( {w_r^i;b} \right)                                                                                                                                                                                            \\
		\mathbf{w}_{r+1}^{i} & =\left\{\begin{array}{ll}\mathbf{v}_{r+1}^{i} & \text { if } r+1\notin \mathcal{I}_{E} \\ \text { samples }  \mathcal{C}_{r}  \text { and average }\left\{\mathbf{v}_{r+1}^{i}\right\} & \text { if } r+1\in  \mathcal{I}_{E}. \end{array}\right.
	\end{aligned}
\end{equation*}

Nex, we have

\begin{equation*}
	\begin{aligned}\left\|\overline{\mathbf{w}}_{r+1}-\mathbf{w}^{*}\right\|^{2} &=\left\|\overline{\mathbf{w}}_{r+1}-\overline{\mathbf{v}}_{r+1}+\overline{\mathbf{v}}_{r+1}-\mathbf{w}^{*}\right\|^{2} \\ &=\underbrace{\left\|\overline{\mathbf{w}}_{r+1}-\overline{\mathbf{v}}_{r+1}\right\|^{2}}_{A_{1}}+\underbrace{\left\|\overline{\mathbf{v}}_{r+1}-\mathbf{w}^{*}\right\|^{2}}_{A_{2}} \\ &+ \underbrace{2\left\langle\overline{\mathbf{w}}_{r+1}-\overline{\mathbf{v}}_{r+1}, \overline{\mathbf{v}}_{r+1}-\mathbf{w}^{*}\right\rangle}_{A_{3}} \end{aligned}
\end{equation*}
Considering unbiasedness of ${\overline w_{r + 1}}$, ${A_3}$ will be vanished when we take expectation over $ \mathcal{C}_{r+1} $.
If $ r+1\notin \mathcal{I}_{E} $, $ A_{1} $ vanishes since $ \overline{\mathbf{w}}_{r+1}=\overline{\mathbf{v}}_{r+1} $. Using \Cref{LEMMA1} to bound ${A_2}$, we get
\begin{equation*}
	\mathbb{E}\left\|\overline{\mathbf{w}}_{r+1}-\mathbf{w}^{*}\right\|^{2} \leq\left(1-\eta_{r} \mu\right) \mathbb{E}\left\|\overline{\mathbf{w}}_{r}-\mathbf{w}^{\star}\right\|^{2}+\eta_{r}^{2} B.
\end{equation*}
When $ r+1\in \mathcal{I}_{E} $, $ A_{1} $, we apply \Cref{LEMMA1} to bound ${A_1}$ and then
\begin{equation}\label{eq-bounda1}
	\begin{aligned} \mathbb{E}\left\|\overline{\mathbf{w}}_{r+1}-\mathbf{w}^{*}\right\|^{2} &=\mathbb{E}\left\|\overline{\mathbf{w}}_{r+1}-\overline{\mathbf{v}}_{r+1}\right\|^{2}+\mathbb{E}\left\|\overline{\mathbf{v}}_{r+1}-\mathbf{w}^{*}\right\|^{2} \\ & \leq\left(1-\eta_{r} \mu\right) \mathbb{E}\left\|\overline{\mathbf{w}}_{r}-\mathbf{w}^{\star}\right\|^{2}+\eta_{r}^{2}(B+C) \end{aligned}
\end{equation}
where \[C\geq \frac{1}{\eta_{r}^{2}} \mathbb{E}_{\mathcal{C}_{r}}\left\|\overline{\mathbf{v}}_{r+1}-\overline{\mathbf{w}}_{r+1}\right\|^{2}.\]

\Cref{eq-bounda1} recursively portrays the distance
between $\overline{\mathbf{w}}_{r+1}$ and
$\mathbf{w}^{*}$,
and we will show how it is a decreasing function.
Let $\Delta_{r} = \mathbb{E}\left\|\overline{\mathbf{w}}_{r}-\mathbf{w}^{*}\right\|^{2}$,
\Cref{eq-bounda1} can be simplified to
\begin{equation}\label{eq-simplify}
	{\Delta _{r + 1}} \leqslant (1 - {\eta _r}\mu ){\Delta _r} + {\eta _r}^2(B + C)
\end{equation}
For decreasing stepsize $ \eta_{r}=\frac{\beta}{r+\gamma} $, we need $ \beta>\frac{1}{\mu} $ and $ \gamma>0$.
$ \eta_{1} \leq \min \left\{\frac{1}{\mu}, \frac{1}{4L}\right\}=\frac{1}{4L} $ and $ \eta_{r} \leq2\eta_{r+E} $
is an important condition for \Cref{eq-simplify}.
Therefore, we can see
\begin{equation}\label{eq-recursion}
	\Delta _{r + 1} \leq \frac{v}{\gamma+r}
\end{equation}
when \[ v=\max \left\{\frac{\beta^{2}(B+C)}{\beta \mu-1},(\gamma+1)\left\|\mathrm{w}_{1}-\mathrm{w}^{*}\right\|^{2}\right\}.\]

Now we use mathematical induction. When $r = 1$, \Cref{eq-recursion} holds from the definition of $v$.
Assuming that the conclusion holds for some $r$,
then at $r+1$,
\begin{equation*}
	\begin{aligned}
		\Delta_{r+1} & \leq\left(1-\eta_{r} \mu\right) \Delta_{r}+\eta_{r}^{2} B \\ & \leq\left(1-\frac{\beta \mu}{r+\gamma}\right) \frac{v}{r+\gamma}+\frac{\beta^{2} B}{(r+\gamma)^{2}} \\ &=\frac{r+\gamma-1}{(r+\gamma)^{2}} v+\left[\frac{\beta^{2} B}{(r+\gamma)^{2}}-\frac{\beta \mu-1}{(r+\gamma)^{2}} v\right] \\ & \leq \frac{v}{r+\gamma+1}
	\end{aligned}
\end{equation*}

Using strong convexity of $ {F(\cdot)} $,
\begin{equation*}
	\mathbb{E}\left[F\left(\overline{\mathbf{w}}_{r}\right)\right]-F^{*} \leq \frac{L}{2} \Delta_{r} \leq \frac{L}{2} \frac{v}{\gamma+r} .
\end{equation*}
When $ \beta=\frac{2}{\mu} $,
$ \gamma=\max \left\{8\frac{L}{\mu}, E\right\}-1$ and with $ \kappa=\frac{L}{\mu} $,
we have $ \eta_{r}=\frac{2}{\mu} \frac{1}{\gamma+r} $,
which satisfies $ \eta_{r} \leq2\eta_{r+E} $.
Therefore,
\begin{equation*}
	\mathbb{E}\left[F\left(\overline{\mathbf{w}}_{r}\right)\right]-F^{*} \leq \frac{\kappa}{\gamma+r}\left(\frac{2(B+C)}{\mu}+\frac{\mu(\gamma+1)}{2}\left\|\mathbf{w}_{1}-\mathbf{w}^{*}\right\|^{2}\right)
\end{equation*}
This proves \Cref{THEOREM1}.


\bibliographystyle{ACM-Reference-Format}
\bibliography{fair-bfl}

\end{document}